\documentclass[a4paper,twocolumn,superscriptaddress,11pt,tightenlines,prl,reprint]{revtex4-1} 

\usepackage[utf8x]{inputenc}
\usepackage{amsmath,amssymb}
\usepackage{bbm}
\usepackage{graphicx}
\newcommand{\ve}{\varepsilon}
\newcommand{\ves}{\varepsilon^{*}}
\newcommand{\ie}{{\it i.e.}}
\newcommand{\cf}{{\it cf.}}
\newcommand{\eg}{{\it e.g.}}

\begin{document}
\title{Propagation failure of excitation waves on trees and random networks}

\author{Nikos E. Kouvaris}
\affiliation{Departament de F\'isica Fonamental, Universitat de Barcelona, Mart\'i i Franqu$\grave{e}$s 1, 08028 Barcelona, Spain}
\email{nikos.kouvaris@ub.edu}
\author{Thomas Isele}
\affiliation{Institut f\"ur Theoretische Physik, Technische Universit\"at Berlin, Hardenbergstra{\ss}e 36, 10623 Berlin, Germany}
\author{Alexander S. Mikhailov}
\affiliation{Department of Physical Chemistry, Fritz Haber Institute of the Max Planck Society, Faradayweg 4-6, 14195 Berlin, Germany}
\author{Eckehard Sch\"oll}
\affiliation{Institut f\"ur Theoretische Physik, Technische Universit\"at Berlin, Hardenbergstra{\ss}e 36, 10623 Berlin, Germany}

%
%
%
%
%
%
%
\begin{abstract}
Excitation waves are studied on trees and random networks of coupled active elements. Undamped propagation of such waves is observed in those networks. It represents an excursion from the resting state and a relaxation back to it for each node. However, the degrees of the nodes influence drastically the dynamics. Excitation propagates more slowly through nodes with larger degrees and beyond some critical degree waves lose their stability and disappear. For regular trees with a fixed branching ratio, the critical degree is determined with an approximate analytical theory which also holds locally for the early stage of excitation spreading in random networks.
\end{abstract}
\maketitle
%
%
%
%
\section{Introduction}
\par
Distributed excitable media are found in a wide range of natural systems including neural cells \cite{izhikevich-book}, heart tissue \cite{keener-book-II} or chemical systems \cite{mik-book-I}. They consist of coupled elements obeying an activator-inhibitor dynamics with a single stable fixed point of rest where small perturbations are damped out. However, a large enough perturbation causes a burst of activity after which the elements return back to their resting state. This results in propagation of an excitation wave. Such media have been broadly studied with continuous reaction-diffusion equations and support a variety of self-organised spatiotemporal patterns like pulses, expanding target waves, or rotating spirals \cite{mik-book-I}. 
\par
Within the last decade, self-organisation of patterns has been considered in networks, where reactions occur on the network's nodes and diffusion is carried out through the links connecting them. Such systems can be formed by diffusively coupled chemical reactors \cite{kar02}, biological cells \cite{big01} or dispersal habitats. The rapid development in network science provides increasing insight on the impact of their architecture upon the emerging collective dynamics \cite{barrat-book-08}. A variety of self-organisation phenomena has been studied in such complex systems including epidemic spreading \cite{COL06,COL08}, synchronisation \cite{BOC06,ARE08} and chimera states \cite{Omelchenko2013}, stationary Turing \cite{nak10} and self-organised oscillatory \cite{HAT13} patterns, as well as pinned fronts \cite{KOU12}. Collective phenomena induced by feedback control  \cite{LEH11,Hata2012,KOU13a} or by noise \cite{Atsumi2013,son13} have also been analysed in networks. 
\par
Recent theoretical and experimental studies in networks of coupled excitable nodes have shown that self-sustained activity \cite{ROX04,tattini12} and spreading of excitation waves \cite{Steele2006} are possible and depend strongly on the network architecture. However, propagation failure of excitation waves, which is a very important aspect in neural and cardiac physiology \cite{keener-book-I,keener-book-II,dockery89,DAH08}, as well as in chemical systems has not yet been systematically analysed in networks.
\par
In this letter we show that propagation failure of excitation waves is significantly influenced by the degree of the nodes. Waves propagate more slowly through nodes with larger degrees and beyond some critical degree they disappear. For regular trees with fixed branching ratio a numerical continuation method could be employed. It reveals that a wave loses its stability and dies out through a saddle-node bifurcation that occurs at the critical degree. For the trees with strong diffusive coupling a kinematical theory \cite{MIK91}, which allows for the analytical determination of the critical degree, could be developed. These approximations hold locally for the early stage of excitation spreading in random networks, where numerical simulations have been performed.
%
%
%
%
\section{Excitable systems on regular trees}
Let us consider a two-component excitable system, where only the activator can diffuse and the inhibitor varies slowly. Such a classical continuous medium is described by,
\begin{eqnarray}
\label{eq:rdcon}
\dot{u}(\mathbf{x},t) &=& f(u,v) +D\nabla^{2} u(\mathbf{x},t)\,,\nonumber\\
\dot{v}(\mathbf{x},t) &=& \ve g(u,v)\,,
\end{eqnarray}
\noindent where $u(\mathbf{x},t)$ and $v(\mathbf{x},t)$ denote local densities of the activator and inhibitor species, respectively. Functions $f(u,v)$ and $g(u,v)$ specify local dynamics of activator and inhibitor, respectively, parameter $\ve$ represents the ratio of their characteristic time scales and $D$ is the diffusion constant. The terms ``activator'' and ``inhibitor'' refer here to the dynamical roles of variables which may have different origins. As an example we choose the FitzHugh-Nagumo (FHN) \cite{izhikevich-book} system,
\begin{equation}
\label{eq:fhn}
f(u,v)=u-\frac{u^{3}}{3}-v\, \ \text{ and } \ g(u,v)=u-\beta\,,
\end{equation}
\noindent where the activator variable $u$ represents fast changes of the electrical potential across the membrane of a neural cell, while the inhibitor variable $v$ has no direct physiological significance; however, it is related to the gating mechanism of the membrane channels. 
\par 
If activator and inhibitor species occupy the nodes of a network and the activator can diffusively be transported over network links to other nodes, then the analog of system (\ref{eq:rdcon}) reads,
\begin{eqnarray}
\label{eq:rdnet}
\dot{u}_i  &=& f(u_i,v_i) +  D\sum_{j=1}^{N}\! T_{ij}(u_j-u_{i})\,,\nonumber\\
\dot{v}_i &=& \ve g(u_i,v_i) \,,
\end{eqnarray}
\noindent where $u_i$ and $v_i$ are the densities of the activator and inhibitor in a network node $i$. The local dynamics on the nodes is described by the functions $f(u_i,v_i)$ and $g(u_i,v_i)$. Diffusional mobility of the activator is taken into account in the summation term, where $T_{ij}$ is the adjacency matrix determining the architecture of the network, whose elements are $1$, if there is a link connecting nodes $i$ and $j$ and $0$ otherwise. Only undirected networks are considered here, {\it i.e.} $T_{ij}=T_{ji}$.  The degree $k_i=\sum_jT_{ji}$ of node $i$ is the number of its connections and plays an essential role in the dynamics of excitation waves. 
\begin{figure}[t!]
\includegraphics[width=0.5\textwidth]{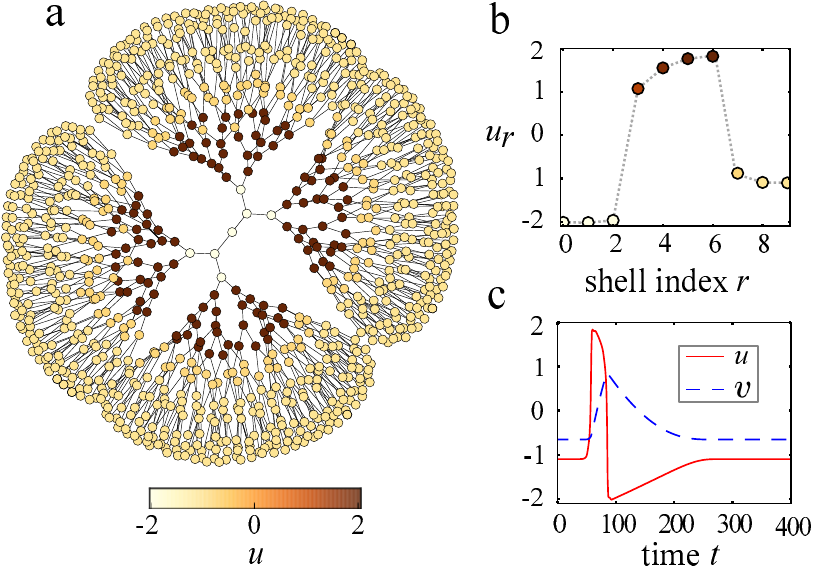}
\caption{(Colour online) Snapshot of an excitation wave which propagates from the root towards the periphery in a regular tree of $10$ shells with branching ratio $2$ and $N=2047$ nodes (a). The wave can also be represented as a pulse by grouping the nodes with the same distance from the root into a single shell (b). The time evolution of a single node (or shell) is shown in (c). Other parameters: $D=0.04,\ \beta=-1.1$, $\ve=0.02$.}
\label{fig:tree}
\end{figure} 
\par
Here we consider a hierarchical organisation of the system (\ref{eq:rdnet}) on a regular tree with  branching ratio $k-1$. In trees, all nodes with the same distance $r$ from the root can be grouped into a single shell \cite{KOU12,KOU13a}. The activator of a node which belongs to the shell $r$ can diffusively be transported to $k-1$ nodes in the next shell $r+1$ and to just one node in the previous shell $r-1$. Introducing the densities $u_r$ and $v_r$ for the activator and the inhibitor in the shell $r$, the evolution of their distribution on the tree can be described by the equations,
\begin{eqnarray}
\label{eq:kchain}
\dot{u}_r  &=& f(u_r,v_r) +  D [u_{r-1}-ku_{r} +(k-1)u_{r+1}]\,,\nonumber\\
\dot{v}_r&=& \ve g(u_r,v_r) \,.
\end{eqnarray}
\noindent  Note that although $k$ is the degree of the nodes and thus can take only integer values, it can be treated as a continuous parameter. In our approximation, instead of investigating propagation of excitation waves directly on a tree network (fig.~\ref{fig:tree} (a)) we study it, using the sequence of coupled shells (fig.~\ref{fig:tree} (b)) described by eqs. (\ref{eq:kchain}). Contrary to chains ($k=2$), where both propagation directions (left or right) of excitation waves are equivalent, propagation from the root to the periphery of a tree ($k>2$) is physically different from the propagation in the opposite direction, i.e. towards the tree root. Here, only excitation waves that start from the root which can -- under appropriate conditions -- propagate towards the periphery are considered.
\par
Excitation waves can be generated by applying a large enough external perturbation to the root of the tree while all other nodes are in the resting state. This perturbation can excite the root. Subsequently, it is possible for excitation to be passed from one node to another, due to the diffusional transport of the activator and it reaches nodes within the same shell at the same time. Thus, propagation of undamped excitation waves, which represent an excursion from the resting state and a relaxation back to it (fig.~\ref{fig:tree} (c)) for each node of the tree, can be supported. A snapshot of such a wave is shown for a particular tree in fig.~\ref{fig:tree}. 
\begin{figure}[t!]
\includegraphics[width=0.491\textwidth]{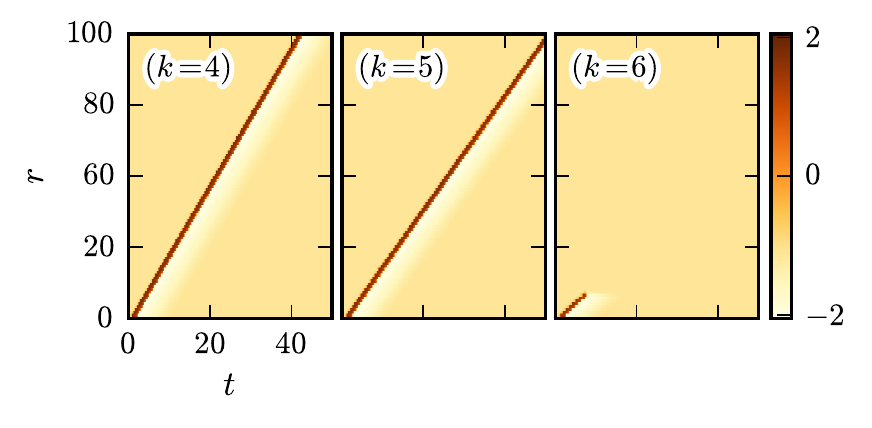}
\caption{(Colour online) Space-time plots of the activator density $u_r$ shows the evolution of an excitation wave in the trees with different node degree $k$. Other parameters as in fig. \ref{fig:tree}.}
\label{fig:spacetime}
\end{figure} 
\par
\begin{figure}[t!]
\includegraphics[width=0.5\textwidth]{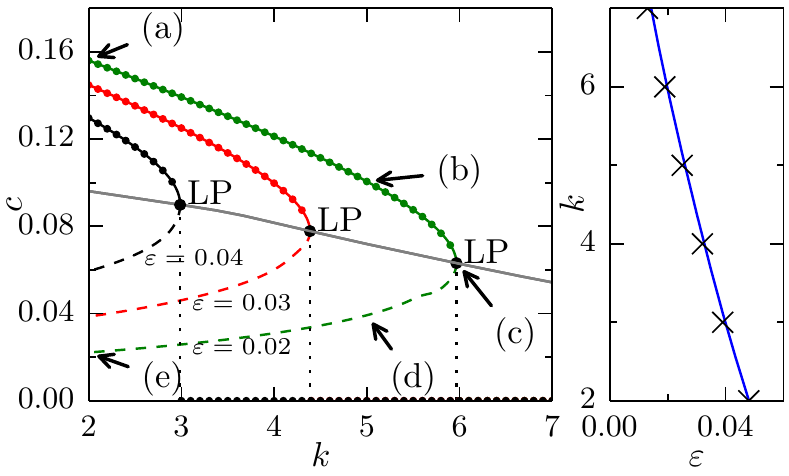}
\caption{(Colour online) (Left) Dependence of the propagation velocity $c$ on the degree $k$ as calculated from numerical simulations (dots) and from numerical continuation (curves) on the ring of 50 shells for different $\ve$. Solid curves show stable solutions while dashed curves show the unstable ones. The letters (a)-(e) denote selected solutions that are shown in fig.~\ref{fig:timeseries}. (Right) The location of the saddle-node bifurcation points (LP) is shown in the ($\ve$,$k$) plane. Other parameters as in fig.~\ref{fig:tree}.}
\label{fig:ck}
\end{figure}

Not all excitable trees can, however, lead to such propagating waves. In fig.~\ref{fig:spacetime} we see for a given set of parameters that excitation waves can propagate from the root towards the periphery in trees with $k=4$ and $k=5$. However, trees with larger branching ratio, \eg{} $k=6$, fail to support the undamped propagation of waves; starting from the root, excitation may be passed to all nodes of some shells with short distance (shortest path length) from the root, but then fails to propagate further and vanishes (see fig.~\ref{fig:spacetime} for $k=6$). Numerical simulations have revealed that excitation waves propagate more slowly in trees with larger $k$ (see fig. \ref{fig:ck}). As $k$ increases, it reaches a critical value where waves propagate with the minimum (positive) velocity. In contrast to bistable trees, where fronts can be pinned or retreated as $k$ becomes larger \cite{KOU12,KOU13a}, excitation waves cannot stop or reverse the direction of their propagation. They become unstable and disappear beyond this critical degree (see fig. \ref{fig:ck}). When $k$ is fixed, the same transition to unstable waves occurs when some critical value of the time scale separation constant $\ve$ is exceeded. The stability analysis of these waves is performed with a numerical continuation method.
%
%
%
%
\par
For this purpose we assume a ring of shells, as described by system (\ref{eq:kchain}). On this ring, each shell is coupled with weight $k-1$ to its neighbour in clockwise direction and with weight $1$ to its neighbour in counterclockwise direction. Locally, such a ring of shells resembles the shells of a tree, globally however, it cannot be mapped to a tree. By locally we mean in this context that the wave-like solutions we are examining are constrained to such a small part of the ring that they do not interact with themselves; or formulated differently, that the fraction of nodes lying in the resting state is always large enough to clearly separate the leading edge of one excitation wave from the refractory phase of the preceding wave. Therefore, in the regimes of parameters where the wave-like solutions are localized within only a small fraction of this ring, we expect to observe the same dynamics as in the shells of an actual tree network.
\par
For continuation purposes this construction has the advantage that a traveling wave on this ring of shells is a periodic orbit of $d\times S$ ordinary differential equations (ODE), where $S$ is the number of shells of the tree and $d$ is the dimension of the local dynamics on each shell ($2$ in the case of the FHN system), Such a system can easily, but possibly at large numerical expense, be continued, using a numerical continuation software, \eg{} AUTO-07p \cite{auto07}. 
\par
\begin{figure}[t!]
\includegraphics[width=0.491\textwidth]{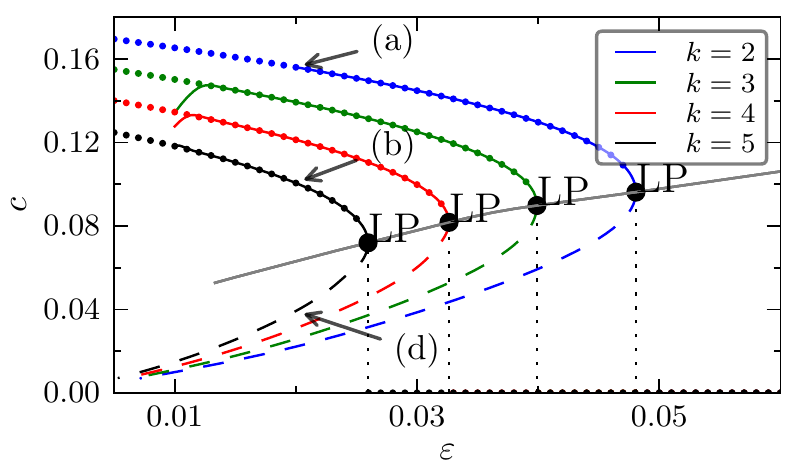}
\caption{(Colour online) Dependence of the propagation velocity $c$ on the parameter $\ve$ as calculated from numerical simulations (dots) and from numerical continuation (curves) on the ring of 50 shells for different $k$. Solid curves show stable solutions while dashed curves show the unstable ones. The letters (a),(b),(d) denote selected solutions that are shown in fig.~\ref{fig:timeseries}. Other parameters as in fig.~\ref{fig:tree}.}
\label{fig:ce}
\end{figure}

We consider such a ring consisting of $S=50$ shells, obeying the FHN dynamics and diffusively coupled with strength $D=0.04$. Thus, we proceed to the continuation of the periodic solutions of a system of $100$ coupled ODEs. From the continuation, we directly obtain the stability properties of the solution, as well as the location of the limit points (LP) of the saddle-node bifurcations, which are marked in fig.~\ref{fig:ck}. We see that the continuation of periodic orbits in the ring of shells gives exactly the same results for the propagation velocity $c$ and for the stability as those obtained from the direct simulation on trees. Clearly, the transition from undamped propagation to unstable excitation waves in trees takes place through a saddle-node bifurcation. Similar dynamical behaviour is observed for a given value of the degree $k$ as $\ve$ increases (see fig.~\ref{fig:ce}). 
\par
\begin{figure}[t!]
\includegraphics[width=0.5\textwidth]{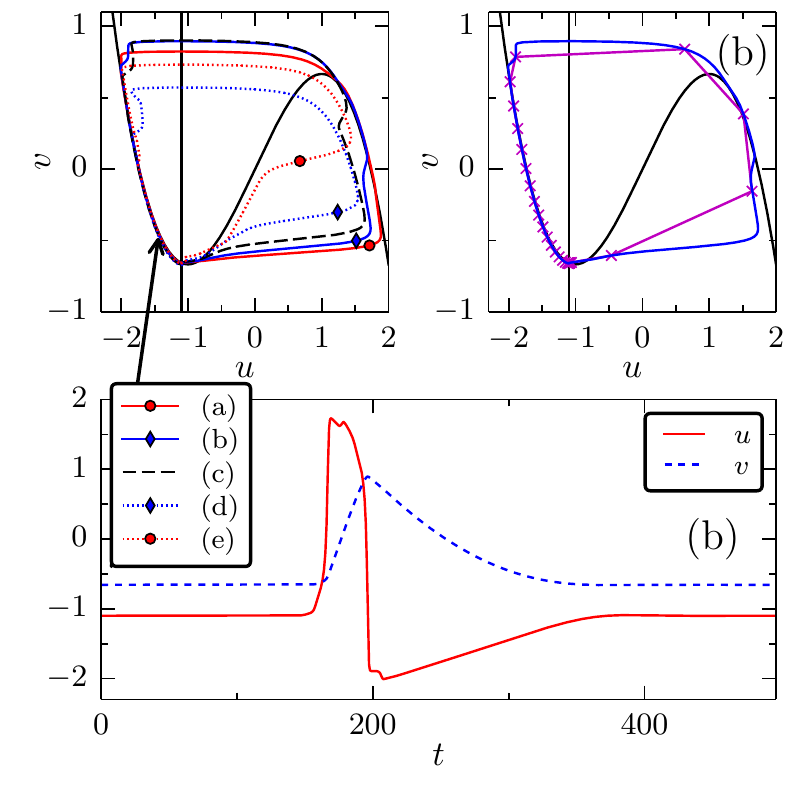}
\caption{(Colour online)
Different solutions of eqs.~(\ref{eq:kchain}) obtained by continuation on the ring of $50$ shells are shown:
(a) stable solution for $k=2$ with $c=0.1561$ ($T=320.349$),
(b) stable solution for $k=5$ with $c=0.1006$ ($T=497.047$),
(c) Limit Point solution for $k\approx5.9766$ with $c=0.0631$ ($T=791.849$),
(d) unstable solution for $k=5$ with $c=0.0392$ ($T=1274.64$) and
(e) unstable solution for $k=2$ with $c=0.0222$ ($T=2256.39$).
Other parameters as in fig.~\ref{fig:tree}. 
In the upper right panel, solution (b) is shown as the path of densities in phase space ($u$,$v$) together with a snapshot of the state of all shells at one instant of time (magenta crosses) and the nullclines (grey) of system (\ref{eq:fhn}), while in the lower panel the evolution of activator (red curve) and inhibitor (blue curve) densities on one shell are shown. The upper left panel shows a comparison of the solutions (a)-(e) as the paths of densities in phase space ($u$,$v$). The location of the solutions is also marked in fig.~\ref{fig:ck}.
Other parameters as in fig.~\ref{fig:tree}. }
\label{fig:timeseries}
\end{figure}

The asymmetric coupling (weight $1$ to the previous and $k-1$ to the next shells) in combination with the discreteness of the system affects also the shape of the excitation waves. As can clearly be seen in fig.~\ref{fig:timeseries}, the trajectory in the ($u$,$v$) plane followed by each shell, and thus the corresponding timeseries of one shell $r$, appears with two ``dips'' (see also supplementary movie1.avi). The reason is that in eqs.~(\ref{eq:kchain}), the diffusive coupling term for a shell $r$ whose current state is already moving on the slow manifold and thus changing on the slow timescale, can vary on the fast timescale, when shell $r+1$ moves on the fast manifold from the rest state to the excited state (see upper right panel in fig.~\ref{fig:timeseries}). Because shell $r+1$ is coupled to $r$ with weight $k-1$, whereas shell $r-1$ is coupled to $r$ with weight 1, the ``force'' exerted by the coupling term, ``pulls'' the shell $r$ in the direction of the resting state, when shell $r+1$ is still close to the resting state. It does so more strongly for larger $k$. 
%
%
%
%
\section{Kinematical theory for excitable trees}
\par
The propagation velocity $c$ is unique in regular trees with fixed branching ratio and depends on the parameters $k,\ \ve,\ D$ and $c_0$; $c_0$ is the velocity of a bistable front in the absence of inhibitor (\cf{} \cite{KOU12}). Here we calculate an analytical expression for the dependence $c=c(k)$ by extending for the trees the kinematical theory proposed by {\it Mikhailov} and {\it Zykov} in \cite{MIK91}.
\par
Let us assume a regular tree with infinite hierarchical levels and strong diffusive coupling (\ie{}, large $D$). In such a tree, we can obtain an approximation for the continuous limit by substituting $u_{r-1}$ and $u_{r+1}$ in eqs.~(\ref{eq:kchain}) with their Taylor expansions $u_{r-1}  \approx u_{r} -\nabla u + \Delta u/2\,$ and $u_{r+1}  \approx u_{r} + \nabla u + \Delta u/2\,$. Then, in the continuous limit, the system (\ref{eq:kchain}) reads,
\begin{eqnarray}
\label{eq:kchain2con}
\dot{u}  &=& f(u,v) +  \frac{Dk}{2} \Delta u + D(k-2) \nabla u\,,\nonumber\\
\dot{v} &=& \ve g(u,v) \,.
\end{eqnarray}
\noindent By introducing the moving reference frame $\xi = r-ct\,$ and by assuming that the profile of the wave is stationary in this frame, we can reduce the system (\ref{eq:kchain2con}) to a system of two ODEs,
\begin{eqnarray}
\label{eq:CRDA}
-[c+D(k-2)] u^{\prime}  &=& f(u,v) +  \frac{kD}{2} u^{\prime\prime}\,,\nonumber\\
-cv^{\prime}  &=& \ve g(u,v) \,,
\end{eqnarray}
\noindent where $u=u(\xi)$ and $v=v(\xi)$; prime denotes a derivative with respect to the moving coordinate $\xi$. Subsequently, if we replace $\ve$ by the modified parameter $\ves$ in the latter equation, where,
\begin{equation}
\label{eq:epsilonstar}
\ves=\ve\left[1+\frac{D(k-2)}{c}\right]\,,
\end{equation}
\noindent we take the system of equations,
\begin{eqnarray}
\label{eq:rdestar}
-[c+D(k-2)] u^{\prime}  &=& f(u,v) +  \frac{kD}{2} u^{\prime\prime}\,,\nonumber\\
-[c+D(k-2)] v^{\prime}  &=& \ves g(u,v) \,,
\end{eqnarray}
\noindent which describes the propagation of an excitation wave from the root towards the periphery in the same tree as system (\ref{eq:CRDA}), but with time scale separation parameter $\ves$ instead of $\ve$. Therefore, the propagation velocity is 
\begin{equation}
\label{eq:vel1}
c^*= c + D(k-2)\,.
\end{equation}
Substitution of $c$ from eq. (\ref{eq:vel1}) into (\ref{eq:epsilonstar}) yields,
\begin{equation}
\label{eq:rewrite}
 c \equiv c(\ve)= \frac{D (k-2)\ve}{\ves-\ve}\,.
\end{equation}
If we know the function $c(\ve)$ we can find the solution of eq. (\ref{eq:rewrite}) \cite{MIK91}. Here we do not have an analytical expression for this function. However, we have found in the numerical simulations that for very small $\ve$ the velocity $c$ depends linearly on this parameter, \ie{},
\begin{equation}
\label{eq:vel0}
c(\ve) = c_0 (1-\chi\ve) \,,
\end{equation}
\noindent where $\chi$ is a numerical factor independent of $\ve$. Substituting expressions for $\ves$ and $c^*$ into eq.~(\ref{eq:vel0}) and solving the resulting equation we find an analytical expression for the velocity,
\begin{eqnarray}
\label{eq:velsols}
c &=& \frac{1}{2} 
\left[c_0 (1- \ve\chi) - D\left(k-2\right)\right] \\
&&\:\pm
\frac{1}{2}\left\{\left[D(k-2)-c_{0}\left(1-\ve\chi\right)\right]^{2} -4c_{0}D\ve\chi\left(k-2\right)\right\}^{1/2}\, \nonumber
\,.
\end{eqnarray}
\par
\begin{figure}[t!]
\includegraphics[width=0.5\textwidth]{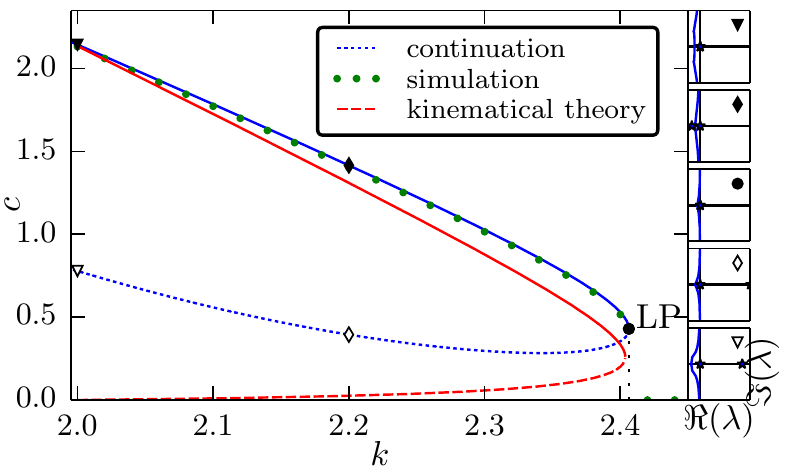}
\caption{(Colour online) Propagation velocity $c$ vs. node degrees $k$. Red curves correspond to the two branches of eq.~(\ref{eq:velsols}), blue curves are obtained by continuation of the eqs.~(\ref{eq:profile_eqs}), green dots have been calculated from the numerical simulations. At the right edge of the figure, the leading part of the spectrum $\lambda \in \mathbbm{C}$ around the wave is shown for stable and unstable solutions at $k=2.0$ ($\blacktriangledown,\triangledown$) and $2.2$ ($\blacklozenge,\lozenge$) as well as for the limit point solution at $k\approx2.406$ ($\bullet$), including the unstable eigenvalue and the zero eigenvalue corresponding to the Goldstone mode of translation invariance. Other parameters: $\ve=0.02,\ D=4,\ c_0=2.178,\ \chi=1.074$.}
\label{fig:vel}
\end{figure}

The two branches of eq.~(\ref{eq:velsols}) are shown in fig.~\ref{fig:vel} (red curves). The lower (dashed) and the upper (solid) branch correspond to the unstable and the stable solution of $c=c(k)$, respectively. As $k$ increases both solutions move toward each other and they merge at the critical value
\begin{equation}
\label{eq:kcr}
k_\text{cr}=\frac{2D + c_{0}\left(1-\sqrt{\ve\chi}\right)^2}{D}\,,
\end{equation}
\noindent where the first derivative of $c(k)$ tends to infinity. No excitation waves can propagate in trees with $k>k_\text{cr}$. Note that at $k_\text{cr}$, waves can still propagate with their minimum critical velocity,
\begin{equation}
\label{eq:vcr}
c_\text{cr}=c_{0}(\sqrt{\ve\chi}-\ve\chi)\,.
\end{equation}
\noindent As we see in fig.~\ref{fig:vel}, our theory allows for a very good estimate of the critical degree $k_\text{cr}$, however, it fails to predict the exact critical velocity $c_\text{cr}$. This disagreement is a consequence of the additional assumption of the linear dependence (\ref{eq:vel0}) of the velocity $c$ of a wave on $\ve$. Such dependence is approximately valid only for small values of this parameter, \ie{} if $\ve\ll 1$. It should hold not only for $\ve$, but also for $\ves$. While $\ve = 0.02$ and thus small, the renormalised parameter $\ves$ increases with $k$ and, at the critical point in fig.~\ref{fig:vel}, it reaches the value $\ves = 0.14$ which is not small enough. The accuracy of the kinematical theory can be further approved (see \cite{MIK91}) by using the actual numerically calculated and nonlinear dependence of the velocity on $\ve$, instead of equation (\ref{eq:vel0}). 
\par
At the continuum limit, we can calculate the velocity and the stability of the excitation waves also by using the profile equations, which are obtained by writing the system (\ref{eq:CRDA}) as a system of first order ODEs,
\begin{eqnarray}
\label{eq:profile_eqs}
u'   &=&  w       \,,                               		\nonumber \\
v'   &=&         - c^{-1} \ve g(u,v)  \,,             	\nonumber \\
w'   &=& -2(Dk)^{-1}\left\{f(u,v) + \left[c+ D (k-2)\right]w \right\}\,.
\end{eqnarray}
\noindent The root of eqs.~(\ref{eq:fhn}) is a fixed point of eqs. (\ref{eq:profile_eqs}) (with $w=0$). An excitation wave on an infinitely extended chain of shells as described by eq.~(\ref{eq:kchain}) appears as a homoclinic trajectory of eqs.~(\ref{eq:profile_eqs}). But such a special trajectory exists only at a definite value of the parameter $c$, which is the propagation velocity of the excitation wave \cite{mik-book-I}. 
\par
Here, instead of numerically continuing these homoclinic trajectories, we continue periodic trajectories, which are very close to the homoclinic ones for large periods. The stability analysis of these periodic trajectories has been performed using Bloch expansion and continuation to calculate the (essential) spectrum which determines the stability of these traveling waves. The method is explained in detail in \cite{RAD07b,SAN02b}. We find that the destabilisation of the excitation waves occurs through a saddle-node bifurcation, where an isolated eigenvalue crosses the imaginary axis exactly at the limit point. The rest of the spectrum is always in the left half-plane, except for one eigenvalue exactly at zero which corresponds to the Goldstone mode of translation invariance. The spectra for selected values of $k$ , including $k\approx2.406$ which corresponds to the limit point solution, are plotted at the right edge of 
fig.~\ref{fig:vel}.
%
%
%
%
\section{Application to random networks}
\begin{figure}[b!]
\includegraphics[width=0.5\textwidth]{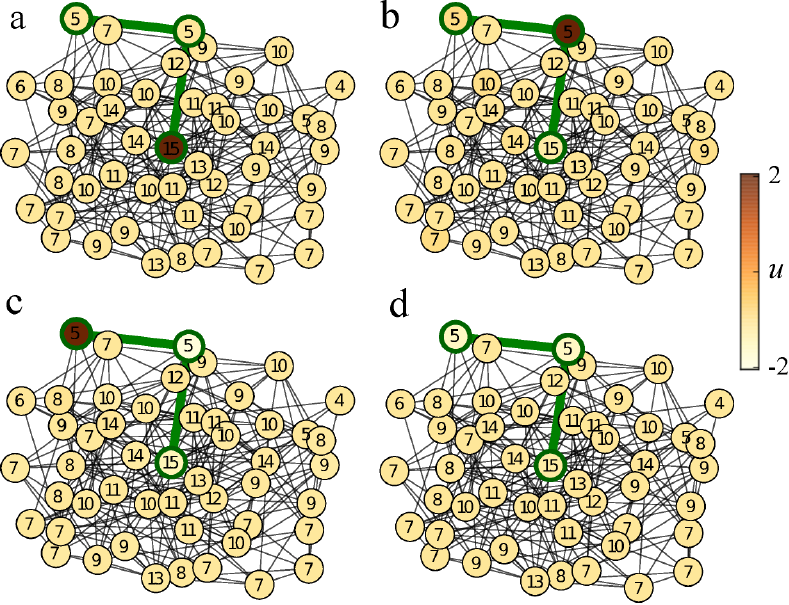}
\caption{(Colour online). Evolution of the activator density for an Erd\"os-R\'enyi random network with $N=50$ nodes and mean degree $\langle k\rangle=10$. Excitation is applied to the hub (a), it consequently propagates to the neighbouring nodes with degrees $k<k_\text{cr}$ (b),(c), and finally dies out before the nodes with degrees $k>k_\text{cr}$ (d). The propagation path is shown with thick green colour. Node labels denote their degrees. Other parameters as in fig.~\ref{fig:tree}.}
\label{fig:er}
\end{figure} 
Propagation failure of excitation waves has also been observed in random networks. Here we provide an example of system (\ref{eq:rdnet}) on an Erd\"os-R\'enyi network, where the hub node is initially set in the excited state as shown in fig.~\ref{fig:er}(a). Consequently, excitation propagates to certain neighbouring nodes (see fig.~\ref{fig:er}(b),(c)) and finally disappears (fig.~\ref{fig:er}(d)). This behaviour can be understood by our approximate theory for the trees, which holds also locally, for the early stage of propagation, in random networks. For the parameters $D=0.04$ and $\ve=0.02$ our theory predicts that excitation waves become unstable at $k_\text{cr}\approx 5.9766$ (see fig.~\ref{fig:ck}). Indeed, we see in fig.~\ref{fig:er} that excitation can propagate only to the neighbouring nodes with degrees $k<6$. Once it reaches a node whose neighbours have degrees $k>6$, it cannot propagate further and disappears. An interesting behaviour that appears in random networks is that excitation may follow certain paths and not others, depending on the degrees of the corresponding nodes. However, if excitation has already spread far from the origin and a large fraction of network nodes have thus become excited, our proposed theory does not hold for random networks.
%
%
%
%
\section{Discussion}
Excitation waves have been analysed in trees and random networks. They can be initiated at the root of excitable tree networks and propagate towards their periphery, representing an excursion from the resting state and relaxation back to it for each node. The propagation velocity decreases in trees with larger degrees until a critical value $k_\text{cr}$. At this critical degree waves are still stable and propagate, however, with their minimum velocity. Once this threshold is exceeded, undamped propagation is not possible. In contrast to the bistable fronts, excitation waves cannot be pinned or reverse their propagation direction. They become unstable and disappear. The approximate theory we have developed for the trees reveals that the destabilisation of the waves takes place through a saddle-node bifurcation which occurs at the critical degree. Same behaviour has been found in trees of given degree, when parameter $\ve$ is increased.
\par
The results of such analysis are also relevant for understanding the early stage of excitation spreading in random networks. When activation is applied to a node and we look locally at its vicinity with its first neighbours, activation propagates only to the nodes with degrees smaller than the critical degree. Depending on the system parameters, excitation may propagate through some nodes or disappear before nodes with larger degrees. This degree heterogeneity in random networks gives rise to the appearance of some preferred paths where excitation can  propagate. In future studies this property should be considered in the design of networks which might adapt their links according to the emerging dynamics in order to drive the excitation through desired paths and nodes.
%
%
%
%
\acknowledgments
Support from DFG in the framework of SFB 910 ``Control of Self-Organizing Nonlinear Systems'' is gratefully acknowledged. NK acknowledges financial support from the LASAGNE project EU/FP7-2012-STREP-318132, Spanish DGICYT Grant No. FIS2012-38266-C02-02 and J. S. Latsis Public Benefit Foundation in Greece.
%
%
%
%
\bibliography{Kouvaris_Isele_Mikhailov_Shoell.bbl}

\end{document}